\documentclass[twocolumn,amsmath,amssymb,prapplied,longbibliography]{revtex4-2}

\usepackage{graphicx}
\usepackage{dcolumn}
\usepackage{bm}
\usepackage{color}
\usepackage{notes2bib}

\begin{document}

\bibliographystyle{naturemag}


\title{Wafer-scale nanofabrication of telecom single-photon emitters in silicon}

\author{M.~Hollenbach$^{1,2}$}
\email[E-mail:~]{m.hollenbach@hzdr.de}
\author{N.~Klingner$^{1}$}
\author{N.~S.~Jagtap$^{1,2}$}
\author{L.~Bischoff$^{1}$}
\author{C.~Fowley$^{1}$}
\author{U.~Kentsch$^{1}$}
\author{G.~Hlawacek$^{1}$}
\author{A.~Erbe$^{1}$}
\author{N.~V.~Abrosimov$^{3}$}
\author{M.~Helm$^{1,2}$}
\author{Y.~Berenc\'en$^{1}$}
\email[E-mail:~]{y.berencen@hzdr.de}
\author{G.~V.~Astakhov$^{1}$}
\email[E-mail:~]{g.astakhov@hzdr.de}

\affiliation{$^1$Helmholtz-Zentrum Dresden-Rossendorf, Institute of Ion Beam Physics and Materials Research, 01328 Dresden, Germany  \\
$^2$Technische Universit\"at Dresden, 01062 Dresden, Germany\\ 
$^3$Leibniz-Institut f\"ur Kristallz\"uchtung (IKZ), 12489 Berlin, Germany
 }

\begin{abstract}
A highly promising route to scale millions of qubits is to use quantum photonic integrated circuits (PICs), where deterministic photon sources, reconfigurable optical elements, and single-photon detectors are monolithically integrated on the same silicon chip. The isolation of single-photon emitters, such as the G centers and W centers, in the optical telecommunication O-band, has recently been
realized in silicon. In all previous cases, however, single-photon emitters were created uncontrollably in random locations, preventing their scalability. Here, we report the controllable fabrication of single G and W centers in silicon wafers using focused ion beams (FIB) with a probability exceeding 50\%. We also implement a scalable, broad-beam implantation protocol compatible with the complementary-metal-oxide-semiconductor (CMOS) technology to fabricate single telecom emitters at desired positions on the nanoscale. Our findings  unlock a clear and easily exploitable pathway for industrial-scale photonic quantum processors with technology nodes below 100~nm.
\end{abstract}
 
\date{\today}

\maketitle


Quantum technologies based on the generation and state manipulation of single photons enable demanding applications \cite{10.1038/nphoton.2009.229, 10.1038/nphoton.2016.186}.  A prime example of this is linear optical quantum computation using boson sampling, which requires only single photons and linear optical components  \cite{10.1038/35051009, 10.1038/nature03347, 10.1126/science.1142892}. The front-runner demonstration is Gaussian boson sampling with 50 single-mode squeezed states \cite{10.1126/science.abe8770}. 
A general-purpose photonic quantum processor can be built using fusing, cluster states, and nonlinear units  \cite{ns, 10.1063/5.0049372}. The latter can be implemented through photon scattering by a two-level quantum system (i.e., a single-photon emitter) coupled to an optical cavity. The state of the art for the deterministic single-photon sources corresponds to boson sampling with 20 photons using quantum dots (QDs) \cite{10.1103/physrevlett.123.250503}. To ensure indistinguishability, the same QD routes several photons into a delay line. Delay lines up to 27~m can be realized on a single silicon chip  \cite{10.1038/ncomms1876}, which allow the interference of about 100 deterministic photons. However, the scalability of millions of qubits is not realistic with this approach. 

Deterministic single-photon sources monolithically integrated with silicon quantum PIC represent a new tool in quantum photonics \cite{10.1364/oe.397377}, complementing heralded probabilistic sources \cite{10.1063/1.4976737} and offering very-large-scale integration (VLSI) \cite{10.1038/s41565-021-00965-6}. The strategic, long-term goal is the implementation of a photonic quantum processor compatible with present-day silicon technology. Most of the necessary components for cryogenic quantum PICs  are available nowadays, including superconducting single-photon detectors \cite{10.1038/ncomms2307}, delay lines  \cite{10.1038/ncomms1876}, modulators \cite{10.1038/s41467-021-21624-3} and phase shifters \cite{10.1038/s41598-017-06736-5}. The practical implementation of this concept has been largely hampered by the lack of controllable fabrication of single-photon emitters in silicon \cite{10.1364/oe.397377, 10.1038/s41928-020-00499-0}. 

\begin{figure*}[t]
\includegraphics[width=.83\textwidth]{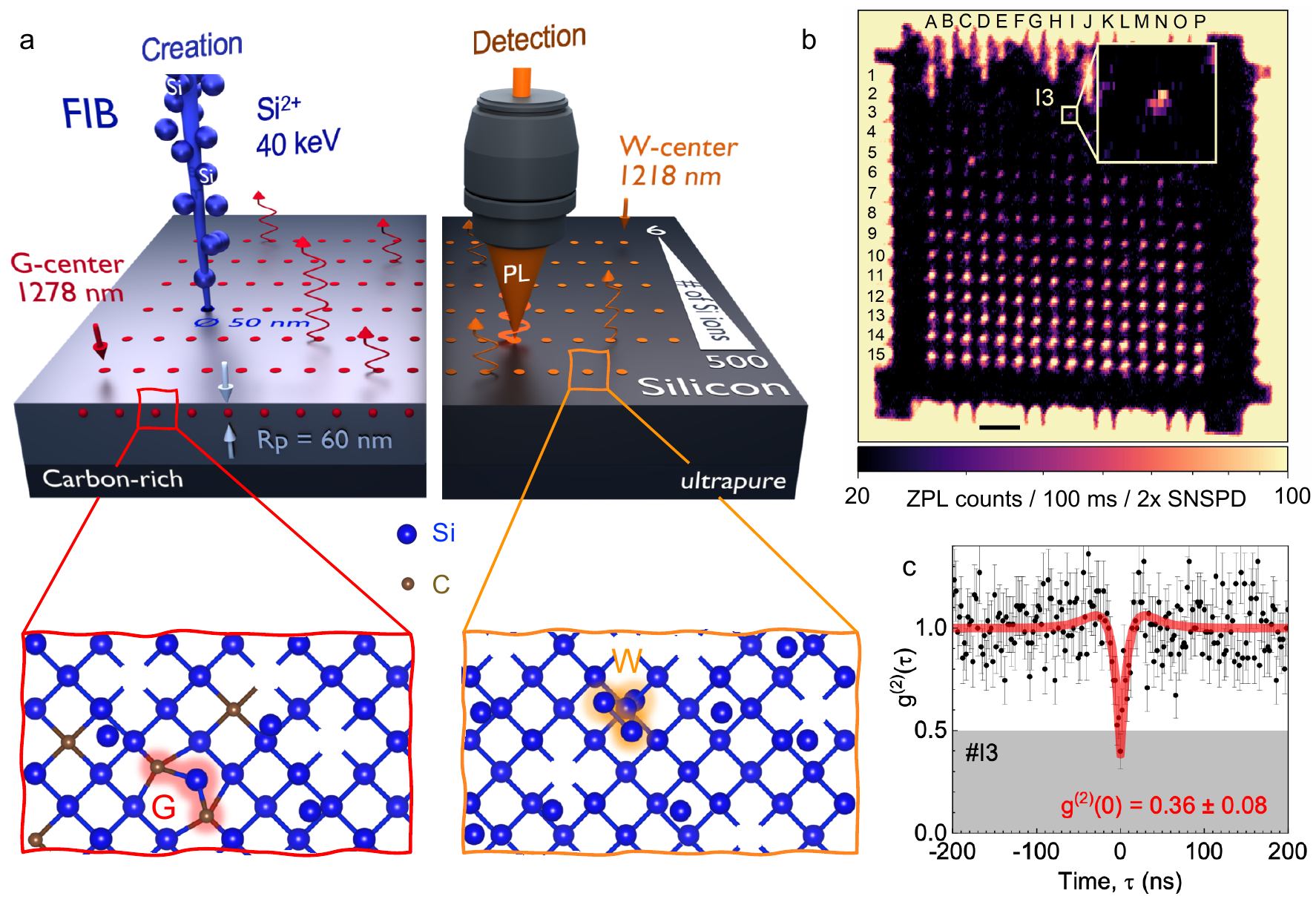}
\caption{Creation and detection of single G and W centers in silicon. a, Schematic of FIB implantation with $\mathrm{Si^{2+}}$ ions and PL collection from single centers. The kinetic energy of $40 \, \mathrm{keV}$ corresponds to an average implantation depth $R_p = 60 \, \mathrm{nm}$. Si implantation into a carbon-rich and an ultrapure silicon wafer results in the formation of the G and W centers, respectively. b, Confocal ZPL ($1278 \, \mathrm{nm}$) intensity map of locally created G centers on an SOI wafer. The number of ions per spot increases logarithmically from nominally 6 (row 1) to 500 (row 15). The pattern frame is created with a fluence $\Phi = 1 \times 10^{11} \, \mathrm{cm^{-2}}$. The scale bar is $20 \, \mathrm{\mu m}$. The inset shows photon emission from a single G-center. The color scale is different from the main panel to increase visibility. c, Second-order autocorrelation function $g^{(2)} (\tau)$ obtained with no BG correction ($\#$I3). The red solid line is a fit to Eq.~(\ref{AntiBunch}) yielding $g^{(2)} (0) = 0.36 \pm 0.06$. }\label{fig1}
\end{figure*}

Recently, a broad variety of single-photon emitters have been isolated in commercial silicon-on-insulator (SOI) wafers \cite{10.1364/oe.397377, 10.1038/s41928-020-00499-0, 10.1103/physrevlett.126.083602, 59a}. This includes single G centers, which are carbon-related color centers emitting in the telecom O-band \cite{10.1364/oe.397377, 10.1038/s41928-020-00499-0}. The atomic configuration of the G center (Fig.~\ref{fig1}a) has been revised several times. According to the latest density functional theory calculations \cite{10.1103/physrevlett.127.196402}, it consists of two substitutional carbon atoms and one interstitial silicon atom in the configuration $\mathrm{C_s-Si_i-C_s}$ distorted from the $\langle 111 \rangle$ bond axis (Fig.~\ref{fig1}a). The spectroscopic fingerprint of the G center is a spectrally narrow zero-phonon line (ZPL) at $\lambda_G =1278 \, \mathrm{nm}$ in the photoluminescence (PL) spectrum \cite{10.1016/0370-1573(89)90064-1}. Another single-photon emitter in silicon is the W center (Fig.~\ref{fig1}a), which is ascribed to a tri-interstitial Si complex $\mathrm{I_3}$ \cite{59a}. Like the aforementioned G-center, it also possesses a single dipole emission, which has been shown to be polarized along the $\langle 111 \rangle$ crystal axis, revealing a ZPL at $\lambda_W =1218 \, \mathrm{nm}$ in the PL spectrum \cite{10.1016/0370-1573(89)90064-1}. 
 
Ensembles of the G and W centers in isotopically purified $\mathrm{^{28}Si}$ crystals reveal extremely narrow linewidths of their ZPLs exceeding the Fourier limit by a factor of two only, which implies marginal spectral diffusion \cite{10.1103/physrevb.98.195201}. This makes the G and W centers very promising candidates for the implementation of spatially separated emitters of indistinguishable photons, where the fine tuning of the emission wavelength can be implemented through the Stark effect or strain control \cite{10.1038/s41534-020-00310-0, 10.1038/s41586-020-2441-3}. 

To date, the protocols for the creation of single-photon emitters in silicon consist of either broad-beam implantation of carbon ions at a low fluence ($\Phi\sim 10^{9} \, \mathrm{cm^{-2}}$) \cite{10.1364/oe.397377} or medium-fluence implantation ($\Phi \sim 10^{12} \, \mathrm{cm^{-2}}$) followed by rapid thermal annealing (RTA) \cite{10.1038/s41928-020-00499-0}. In both approaches, the process of creating single-photon emitters is not controllable, resulting in emitters created at random locations. This poses a major obstacle to the realization of wafer-scale quantum PICs with monolithically integrated and on-demand single-photon sources at desired locations. 

Here, we use a focused ion beam (FIB) \cite{10.1021/acs.nanolett.6b05395, 10.1088/1361-6463/aad0ec, 10.1002/adma.202103235, 10.1021/acs.nanolett.1c04646} to create single G and W centers with nanometer precision. This concept is illustrated in Fig.~\ref{fig1}a. Confirmed by the PL spectra, we unambiguously find that in case of carbon-rich Si wafers, the Si implantation results in the preferable formation of G centers (the left side of Fig.~\ref{fig1}a). For ultrapure silicon wafers and a larger number of Si ions per implantation spot, interstitial complexes rather than G centers are formed, among which are the optically active W centers (the right side of Fig.~\ref{fig1}a). In addition to that, we demonstrate large-scale, CMOS-compatible fabrication of single G centers using broad-beam Si implantation through lithographically defined nanoholes \cite{10.1063/1.4892971}.   

\section*{Creation of single G centers on the nanoscale} 

\begin{figure*}[t]
\includegraphics[width=.93\textwidth]{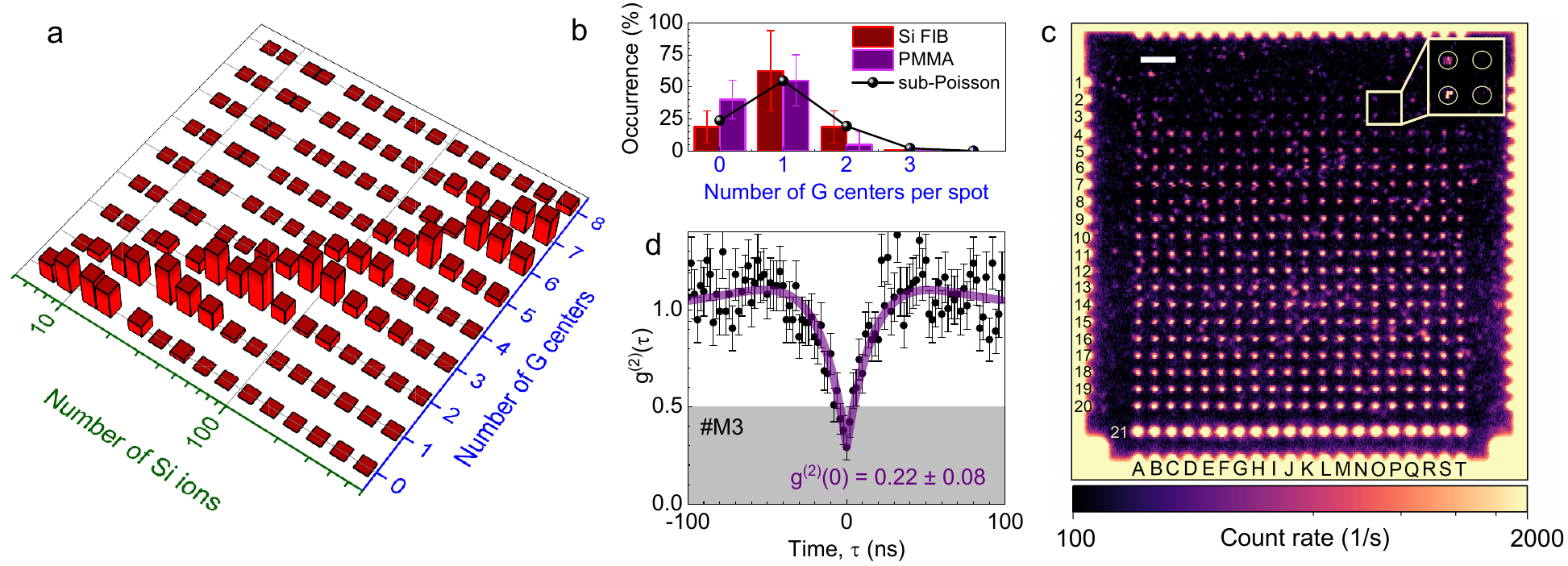}
\caption{Scalable fabrication of single G centers with sub-100-nm precision in an SOI wafer. a, Statistics histogram representing the probability distribution of the G centers depending on the number of implanted Si ions. b, The occurrence probability of G centers for FIB implantation (on average $25$ Si ions per spot) and Si broad-beam implantation (fluence $1 \times 10^{12} \, \mathrm{cm^{-2}}$) through PMMA holes (nominal diameter $40 \, \mathrm{nm}$). The solid line represents the sub-Poisson distribution with $\mu = 4$ as described in the text.  c, Confocal PL intensity map of locally created G centers in an SOI wafer through a PMMA mask using broad-beam Si implantation. The nominal hole diameter increases from $30 \, \mathrm{nm}$ (row 1) to $400 \, \mathrm{nm}$ (row 20). The PL is collected using a BP filter $\Delta \lambda = 50 \, \mathrm{nm}$ at $\lambda =1275 \, \mathrm{nm}$. The scale bar is $20 \, \mathrm{\mu m}$. The inset shows four implanted spots with single and no G centers. d, Second-order autocorrelation function $g^{(2)} (\tau)$ obtained  with no  BG correction ($\#$M3). The purple solid line is a fit to Eq.~(\ref{AntiBunch}) yielding $g^{(2)} (0) = 0.22 \pm 0.08$. }\label{fig3}
\end{figure*}

To create G centers in a commercial SOI wafer (IceMOS tech.), we perform FIB implantation with double-charged  $\mathrm{Si^{2+}}$ ions (Fig.~\ref{fig1}a). The residual carbon concentration is estimated to be in the range of $10^{16} \, \mathrm{cm^{-3}}$ \cite{10.1364/oe.397377}. The Si ions with a kinetic energy of $40 \, \mathrm{keV}$ are focused to a spot size of about $50 \, \mathrm{nm}$. Using the Stopping and Range of Ions in Matter (SRIM) software \cite{10.1016/j.nimb.2010.02.091}, we calculate the lateral straggling to be $\pm 25\, \mathrm{nm}$ and the mean implantation depth to be $R_p = 60\, \mathrm{nm}$. The overall spatial resolution is better than  $100\, \mathrm{nm}$, both in-depth and laterally (Supplementary Fig.~S1). 

We generate a FIB pattern consisting of a frame with a dimension of $200 \times 200 \, \mathrm{\mu m}^2$ and $15 \times 16$ individual spots.
The frame is created by implanting Si ions at a fluence $\Phi = 1 \times 10^{11} \, \mathrm{cm^{-2}}$. The average number of implanted Si ions per spot is the same in each row and increases logarithmically from  $\overline{n}_{\mathrm{Si}} =6$ Si ions for the row 1 to $\overline{n}_{\mathrm{Si}} = 500$ Si ions for the row 15. A detailed list of the averaged number of implanted Si ions ($\overline{n}_{\mathrm{Si}}$) per spot is given in Supplementary Table~SI. We use the chess notation to label each implanted spot. 

After creating the FIB pattern, the samples are measured in a home-built confocal scanning microscope at $T = 6.3 \, \mathrm{K}$ under a continuous wave (cw) laser excitation at 637 nm (Supplementary Fig.~S2).  Figure~\ref{fig1}b shows a confocal ZPL map. To attenuate the background (BG) contribution, which may be related to the presence of defect states in the bandgap, we use a long pass (LP) filter ($\lambda > 1250 \, \mathrm{nm}$) in combination with a narrow bandpass (BP) filter ($\Delta \lambda = 1 \, \mathrm{nm}$) whose central wavelength coincides with the ZPL of the G center $\lambda_G =1278 \, \mathrm{nm}$.

To determine the number of G centers in the implanted spots, we measured the second-order autocorrelation function $g^{(2)} (\tau)$ using Hanbury-Brown-Twiss interferometry (Supplementary Fig.~S2). The collected photons are coupled to a single-mode fiber and split with a 50/50 fiber beam splitter. The photons are then detected with two superconducting-nanowire single-photon detectors (SNSPDs) with an efficiency  $> 90 \%$ in the telecom O-band. The photon detection statistics are recorded with a time-tagging device. An example of such a second-order autocorrelation function from spot $\#$I3 is shown in Fig.~\ref{fig1}c with no BG corrections. It is fitted to  \cite{10.1038/ncomms8578}
\begin{equation}
 g^{(2)} (\tau) =  \frac{N-1}{N} +  \frac{1}{N} \left[ 1 - (1 + a) e^{-| \frac{\tau}{\tau_1} | } + a  e^{- | \frac{ \tau }{\tau_2} | }\right].
 \label{AntiBunch}
\end{equation}
Here, $N$ corresponds to the number of single-photon emitters. The fit to Eq.~(\ref{AntiBunch}) yields $g^{(2)} (0) = 0.36 \pm 0.08 < 0.5$ ($N < 2$), which unambiguously points to a single G center. Remarkably, this G center demonstrates stable operation over hours, with no indication of instability of either the ZPL or the spectrally integrated photon count rate  (Supplementary Fig.~S3). 

To increase the photon count rate and consequently decrease the recording time of $g^{(2)} (\tau)$ in Fig.~\ref{fig1}c,  we use a BP filter with  $\Delta \lambda = 50 \, \mathrm{nm}$ at $\lambda =1275 \, \mathrm{nm}$ instead of the narrow-band filter as in Fig.~\ref{fig1}b. This results in an additional BG contribution to the signal. The autocorrelation function can be corrected due to the presence of the BG as \cite{10.1364/ol.25.001294}
\begin{equation}
 g_{corr}^{(2)} (\tau) = \frac{g^{(2)} (\tau) - (1 - \rho^2)}{\rho^2} .
 \label{Correction}
\end{equation}
The constant factor $\rho = (I-B)/I$ takes into account the count rate from an implanted spot ($I$) and the BG, i.e., the count rate from the location in the immediate surrounding the implanted spots ($B$). After this correction, we obtain  $g_{corr}^{(2)} (0) \approx 0$ for spot $\#$I3. Using this approach, we determine the number of single G centers in other implanted spots (Supplementary Fig.~S3).

\section*{Fabrication statistics} 

The photon count rate from the implanted spots reveals step-like changes rather than continuous variation. We assume that the count rate is proportional to the number of color centers  \cite{10.1063/1.4892971} per implantation spot. To estimate an average count rate from the single G center, we use 
\begin{equation}
 I_G =\frac{\sum_i (I_i - B) }{\sum_i N_i} .
 \label{Calibration}
\end{equation}
Here, $I_i$ is the count rate at spot $i$ in Fig.~\ref{fig1}b obtained from a Gaussian fit (Supplementary Fig.~S2) and  $N_i$ is the number of G centers established from the BG-corrected autocorrelation function following Eq.~(\ref{Correction}). We then estimate the number of the G centers in all implanted spots as $N_i = \mathrm{round} \left[(I_i-B)/I_G\right]$. 

Figure~\ref{fig3}a summarizes the statistical distribution of the number of G centers ($N$) depending on the average number of implanted Si ions ($\overline{n}_{\mathrm{Si}}$).  
The  mean value of $N$ increases with $\overline{n}_{\mathrm{Si}}$ following a sublinear dependence as expected \cite{10.1364/oe.397377}. According to the statistics histogram of Fig.~\ref{fig3}a, the optimal number of Si  ions required to create a single G center is $\overline{n}_{\mathrm{Si}} = 25$ (row 5). The occurrence probability for a different number of G centers, in this case, is presented in Fig.~\ref{fig3}b (the red histogram data Si FIB). The probability to create a single G center is as high as  $(62 \pm 31)\%$, while there is a lower but non-zero probability of creating multiple or no G centers at the implantation spots. Though within the error bars the distributions of Fig.~\ref{fig3}a and b can be described by the Poisson function, there is a strong indication that the experimental data deviate from it. Considering that the G center is a composite defect consisting of three atoms, we can reproduce the sub-Poisson statistics shown by the solid line in Fig.~\ref{fig3}b (Supplementary Fig.~S4).

\begin{figure}[t]
\includegraphics[width=.48\textwidth]{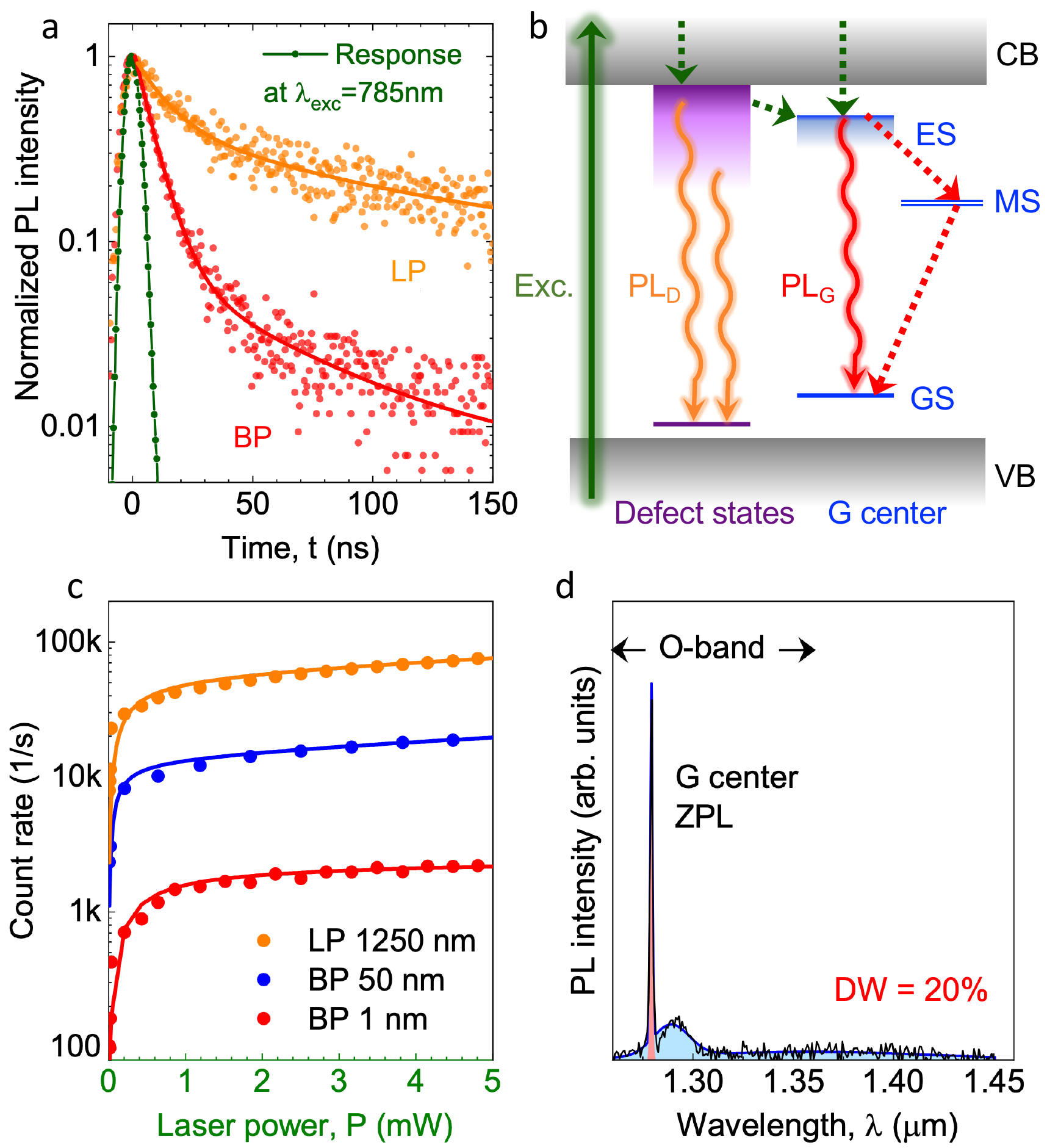}
\caption{Photoexcitation of G centers.  a, PL decay of the locally created G centers obtained with a LP filter $\lambda > 1250 \, \mathrm{nm}$ (orange) and a BP filter $\Delta \lambda = 1 \, \mathrm{nm}$ at $\lambda_G =1278 \, \mathrm{nm}$, corresponding to the ZPL of the G center (red). The solid lines are fits to a bi-exponential decay.  The  excitation laser pulse at $\lambda_{exc} = 785 \, \mathrm{nm}$ is also shown for comparison (green). b, Schematic representation of the excitation and recombination processes of the G center ($\mathrm{PL_G}$)  in the vicinity of bandgap defect states ($\mathrm{PL_D}$). c, Count rate of a single G center as a function of the excitation power in the presence of BG obtained with different optical filters:  LP filter $\lambda > 1250 \, \mathrm{nm}$ (orange), BP filter $\Delta \lambda = 50 \, \mathrm{nm}$ at $\lambda =1275 \, \mathrm{nm}$  (blue) and  BP filter $\Delta \lambda = 1 \, \mathrm{nm}$ at $\lambda_G =1278 \, \mathrm{nm}$ (red). The solid lines are fits to Eq.~(\ref{Power}). d, PL spectrum of a single G center, obtained at $P = 100 \, \mathrm{\mu W}$. A multi-Gauss fit over the ZPL and PSB's (blue solid line) yields a Debye-Waller factor $\mathrm{DW} = 20\%$. } \label{fig2}
\end{figure}

To analyze the BG contribution, we perform time-resolved PL measurements with a LP and a narrow BP filter (Fig.~\ref{fig2}a). The PL spectrum together with the filter transmission wavelengths is shown in Supplementary Fig.~S5. The PL decay is fitted to a bi-exponential function. The fast PL decay with a time constant of about $10 \, \mathrm{ns}$ dominates when the narrow BP filter is tuned to the ZPL \cite{10.1103/physrevb.97.035303}. Therefore, this is associated with the G center. For the spectrally integrated decay, i.e., with the LP filter only, there is a slow contribution with a time constant of about  $70 \, \mathrm{ns}$. This is ascribed to the presence of defect states in the bandgap, which are created during the fabrication of the SOI wafer. The excitation and recombination processes involving the defect states and G centers are schematically presented in Fig.~\ref{fig2}b. This explanation is also confirmed by the excitation power ($P$) dependence of the PL count rate ($I$) for three different filter configurations (Fig.~\ref{fig2}c). It is fitted to 
\begin{equation}
 I (P) =    \frac{I_{G} (\lambda)}{1 + P_0 / P}  +  S_D(\lambda) P\,,
 \label{Power}
\end{equation}
where $I_{G} (\lambda)$ is the saturation count rate and $S_D(\lambda)$ is a spectrally-dependent slope describing the BG contribution. 
The fit of $I (P)$ integrated over the ZPL and the phonon side band (PSB), i.e., with a BP filter $50 \, \mathrm{nm}$, gives $I_{G}  = 13 \times 10^{3}$  counts per second. We find the saturation excitation power for this case $P_0 = 110 \, \mathrm{\mu W}$, which can be reduced using an optimum excitation wavelength according to the PL excitation spectrum \cite{10.1103/physrevb.97.035303, 10.1038/s41928-020-00499-0} (Supplementary Fig.~S5).

\section*{Wafer-scale fabrication of single G centers} 

\begin{figure*}[t]
\includegraphics[width=.93\textwidth]{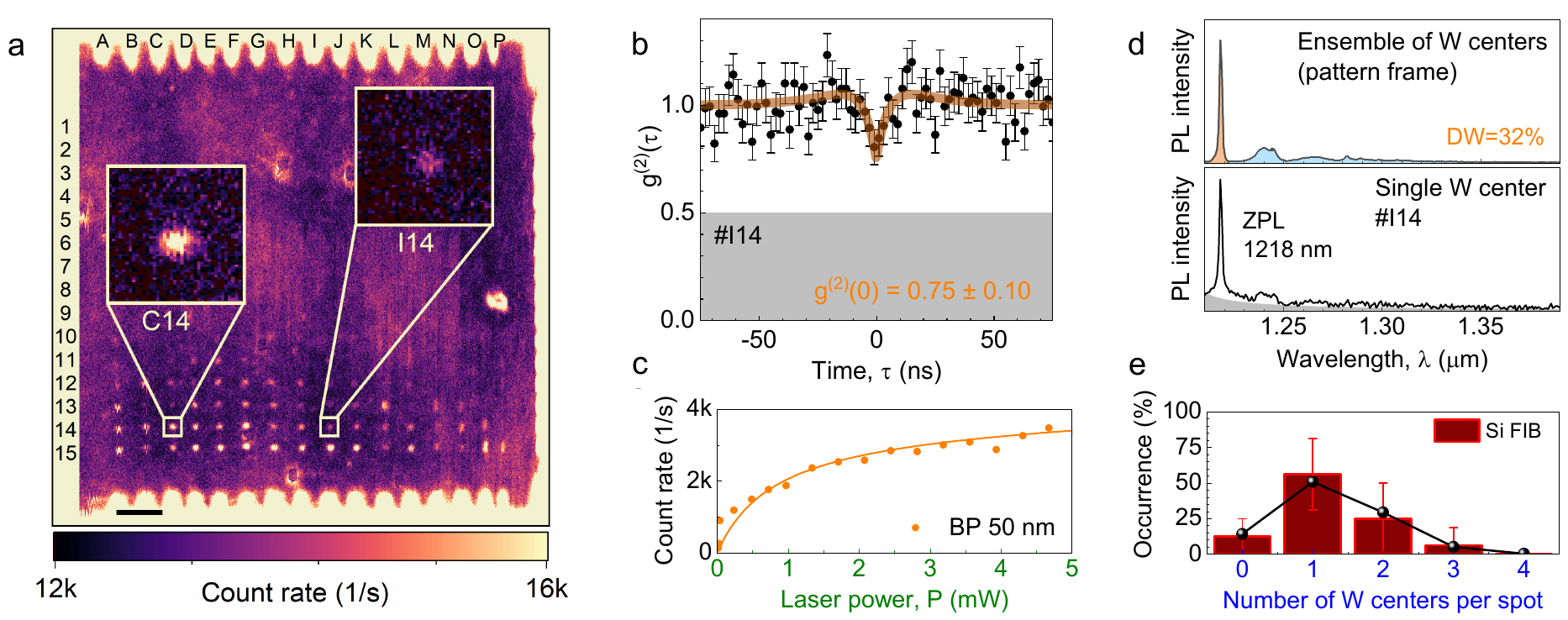}
\caption{Single W centers in ultrapure silicon. a, Confocal PL intensity map of locally created W centers. 
The pattern frame is created with a fluence $\Phi = 1 \times 10^{11} \, \mathrm{cm^{-2}}$. The PL is collected using a BP filter $\Delta \lambda = 50 \, \mathrm{nm}$ at $\lambda =1225 \, \mathrm{nm}$. The scale bar is $20 \, \mathrm{\mu m}$. The insets show the PL from two W centers ($\#$C14) and a single W center ($\#$I14).  
b, Second-order autocorrelation function $g^{(2)} (\tau)$ obtained at the spot $\#$I14 with no BG correction. The BG correction gives $g_{corr}^{(2)} (0) = 0.13^{+0.35}_{-0.13}$. c, Count rate of a single W center after BG subtraction as a function of the excitation power, which is measured with a BP filter $\Delta \lambda = 50 \, \mathrm{nm}$ at $\lambda =1225 \, \mathrm{nm}$. The solid line is a fit to Eq.~(\ref{Power}). d, PL spectrum from the frame and a single W center ($\#$I14). Integration over the ZPL and PSB yields a Debye-Waller factor $\mathrm{DW} = 32\%$. The BG contribution is schematically shown by the shaded area.  e, The occurrence probability of W centers for FIB implantation with on average $384$ Si ions per spot. The solid line represents the sub-Poisson distribution with $\mu = 4.8$ as described in the text. }\label{fig4}
\end{figure*}

To reduce the BG in our commercial SOI wafers, a series of  RTA and furnace annealing (FA) experiments were performed (Supplementary Fig.~S6). We find that the most efficient reduction is obtained with RTA processing at $1000\mathrm{^{\circ}C}$  for $60\, \mathrm{s}$. After optimizing the implantation and annealing parameters, we demonstrate the controllable creation of single G centers using a CMOS-compatible protocol. We first fabricate a PMMA mask with lithographically defined arrays of nanoholes (Supplementary Fig.~S1) having different diameters (Supplementary Table~SII).  Then, we perform a broad-beam implantation with Si ions at a fluence $\Phi = 1 \times 10^{11} \, \mathrm{cm^{-2}}$ and with the same kinetic energy of $40 \, \mathrm{keV}$ as in the FIB experiments.    

A confocal map of the G centers created in $20 \times 20$ nanoholes is depicted in Fig.~\ref{fig3}c. The PL count rate is spectrally integrated over the ZPL and PSBs. As an illustration, we show the autocorrelation function recorded at the spot $\#$M3 with no BG correction (Fig.~\ref{fig3}d).  The fit to Eq.~(\ref{AntiBunch}) yields $g^{(2)} (0) = 0.22 \pm 0.08$  pointing to a single photon emission. Some other $g^{(2)} (\tau)$ measurements of single G centers at different implanted spots are shown in Supplementary Fig.~S7. Based on the $g^{(2)} (\tau)$ measurements and calibrated count rate, we find that more than 50\% of the nanoholes with nominal diameters of $35 \, \mathrm{nm}$ and $40 \, \mathrm{nm}$ (rows  2 and 3, respectively) contain single G centers (Fig.~\ref{fig3}b).

Figure~\ref{fig2}d shows a PL spectrum from the spot with a nominal diameter of $40 \, \mathrm{nm}$ ($\#$M3). It consists of the ZPL at $\lambda_G =1278 \, \mathrm{nm}$ and the PSB with a maximum at around $1290 \, \mathrm{nm}$ \cite{10.1038/s41928-020-00499-0}. The Debye-Waller (DW), i.e., the probability to coherently emit into the ZPL, is an important characteristic of single-photon emitters for their applications in photonic quantum technologies. We find $\mathrm{DW} = 20 \%$, which is the largest value reported to date for individual G centers.

\section*{Creation of single W centers on the nanoscale} 

Finally, we turn to the controlled creation of W center emitters with the ZPL at $\lambda_W =1218 \, \mathrm{nm}$. In oder to locally write W centers,  we use the same procedure as for G-centers in SOI,  with the difference that the substrates are now ultrapure Si wafers with negligible carbon content (Fig.~\ref{fig1}a). After implantation, the sample is annealed at $225\mathrm{^{\circ}C}$ for $300\, \mathrm{s}$ \cite{10.1103/physrevb.98.195201, 10.1364/oe.386450}. Figure~\ref{fig4}a shows a confocal PL map of this pattern. A 50-nm BP filter at $1225 \, \mathrm{nm}$  nm is used to selectively collect the PL emission from the ZPL and the first PSB of W centers.  We optically resolve all the implanted spots in row 15 (on average $500$ Si ions per spot) down to only a few implanted spots in row 10 (on average $113$ Si ions per spot). 

We show an autocorrelation measurement at a spot irradiated with on average $384$ Si  ions ($\#$I14) with no BG correction (Fig.~\ref{fig4}b). The dip at $\tau = 0$ indicates a countable number of W centers ($N \leqslant 5$).  We observe a relatively high BG in the quantum regime (Supplementary Fig.~S8). A possible reason is that we use an established annealing protocol optimized for a dense ensemble of W centers  \cite{10.1364/oe.386450}, which might be not optimum for the creation of individual W centers. Applying the BG correction procedure of Eq.~(\ref{Correction}), we obtain $g_{corr}^{(2)} (0) = 0.13^{+0.35}_{-0.13}$, which indicates that, in fact, we have single-photon emission from this spot. To find the power dependence of the photon count rate from a single W center of  Fig.~\ref{fig4}c, we subtract the BG contribution taken from the non-implanted area between the nearest spots. A fit to Eq.~(\ref{Power}) gives $I_{W} = 3600$ counts per second and $P_0 = 810 \, \mathrm{\mu W}$ (Supplementary Fig.~S8), which is lower than the saturation count rate of the G centers. 

A PL spectrum from a single W center is shown in the lower panel of Fig.~\ref{fig4}d, which is similar to the PL spectrum of an ensemble of W centers (upper panel of Fig.~\ref{fig4}d). We find a $\mathrm{DW} = 32 \%$, which is significantly larger than that for the G center. For low excitation powers ($P \ll P_0$), the PL spectrum and photon count rate remain stable over one day of operation (Supplementary Fig.~S8). For high excitation powers ($P > P_0$), we observe blinking of the ZPL. The origin of this optical instability is beyond the scope of this work. 

Two spots with implantation $\overline{n}_{\mathrm{Si}} = 384$ (row 14)  show a difference in count rate, after the BG correction, of a factor of two, indicating that one contains a single center ($\#$I14) and one contains two single centers ($\#$C14). This is in agreement with the corrected $g_{corr}^{(2)} (0) = 0.52 \pm 0.15$ indicating two-photon emission (Supplementary Fig.~S9). Based on the $g^{(2)} (\tau)$ and the photon count rate analysis of the implanted row 14, we find that, in a similar way to Fig.~\ref{fig3}b, the creation probability of a single W center is $(56 \pm 28)\%$ (Fig.~\ref{fig4}e). Thus, the analysis indicates that the W centers are created with sub-Poisson statistics, as explained in Supplementary Fig.~S4.

\section*{Conclusions} 
In summary, we unambiguously demonstrate the controllable creation of quantum telecom emitters based on single silicon-interstitial- and carbon-related color centers in silicon wafers. These single-photon emitters are created with a spatial resolution better than $100 \, \mathrm{nm}$ and a probability exceeding $50 \%$. Using broad-beam implantation through lithographically defined nanoholes, we demonstrate the wafer-scale nanofabrication of telecom single-photon emitters compatible with CMOS technology for VLSI.  Our results enable the direct realization of quantum PICs with monolithically integrated single-photon sources with electrical control  \cite{10.1364/oe.397377}.  These findings also provide a route for the quasi-deterministic creation of single G and W centers at desired locations of photonic structures \cite{iwa}, tunable cavities \cite{10.1038/s41377-019-0145-y} and SOI waveguides \cite{b9a}. Furthermore, our approach can potentially be applied for the controllable creation of other color centers in silicon, including T centers with optically-interfaced spins \cite{i05}.

\section*{Methods} 

\subsection{Samples}

Two different sets of p-type silicon wafers are utilized for the experiments.  In the case of G centers, we performed our experiments on a commercially available Czochralski (CZ)-grown $\langle 110\rangle $-oriented SOI wafer purchased from IceMOS Technology. This wafer consists of a 12-$\mathrm{\mu}$m-thick Si device layer separated by a 1-$\mu$m-thick silicon dioxide ($\mathrm{SiO_{2}}$) layer from the bulk silicon substrate. The double-side polished 315-$\mu$m-thick substrate is cleaved into $5 \times 5 \, \mathrm{mm^2}$ pieces. The as-grown concentration of carbon impurities for this type of wafers is specified to be higher than $10^{16} \, \mathrm{cm^{-3}}$ \cite{10.1364/oe.397377}. To decrease the natural background contribution, we perform either FA or RTA in a $\mathrm{N_{2}}$ atmosphere. 

To investigate W centers, we use $\langle 100\rangle $-oriented single-side polished, 525-$\mathrm{\mu m}$-thick, ultrapure silicon substrates grown by the float zone (FZ) technique. The residual concentration of carbon and oxygen impurities is less than $5 \times 10^{14} \, \mathrm{cm^{-3}}$ and $1 \times10^{14} \, \mathrm{cm^{-3}}$, respectively, whereas the concentration of boron and phosphorous dopants falls below $7 \times 10^{12} \, \mathrm{cm^{-3}}$. To create the optically active W center, we 
performed FA  at $225^\circ\text{C}$ for $300\, \mathrm{s}$ in $\mathrm{N_{2}}$  atmosphere following fabrication protocols optimized for an ensemble of W centers \cite{10.1103/physrevb.98.195201, 10.1364/oe.386450}. 

\subsection{FIB implantation}\label{FIB} 

We used a customized Orsay Physics CANION Z31Mplus FIB system with a liquid metal alloy ion source (LMAIS). The FIB system is equipped with an in-house-fabricated $\mathrm{Au_{82}Si_{18}}$ ion source, which provides a focused ion beam with a diameter of roughly $50 \, \mathrm{nm}$ \cite{10.1063/1.4947095}. The small focal spot of the FIB offers fast, flexible, maskless and spatially resolved targeted positioning of the implanted ions at the nanoscale. Additionally, the system is equipped with an Wien ExB mass filter to block different ion species and charge states emerging from the ion source. The double-charged $\mathrm{Si^{2+}}$ ions with a nominal beam current between $1$ and $2.5 \, \mathrm{pA}$ have a kinetic energy of $40 \, \mathrm{keV}$ (at $20 \, \mathrm{kV}$ acceleration potential). 

For the FIB implantation of single G and W centers, a custom patterning file is created for both the frame and the single dot arrays, respectively. The frame is implanted with a constant fluence $\Phi \sim 10^{11} \, \mathrm{cm^{-2}}$ to intentionally create a dense ensemble of color centers for the reference and alignment purposes. For the individual single dot arrays with $15 \times 16$ irradiation spots (vertical and horizontal spacing  $10 \, \mathrm{\mu m}$), the number of ions per spot is targeted to be between 6 to 570 with logarithmic incremental steps. The implanted number of Si ions is controlled by the dwell time, such that the desired dose of Si ions is reached.  

\subsection{Broad-beam implantation}

SOI wafers are processed using an RTA at  $1000^\circ\text{C}$ for $3 \, \mathrm{min}$ under $\mathrm{N_{2}}$ atmosphere, $15 \, \mathrm{min}$ of piranha (3 parts $\mathrm{H_2SO_4}$ : 1 part $\mathrm{H_2O_2}$) cleaning is performed to remove residual carbon- and oxygen-terminate the sample surface. Prior to resist spin coating, the samples are ultrasonically cleaned in acetone, rinsed in IPA and blown dry with N2. Next, a layer of positive micro resist (PMMA, 950K A6) with a nominal thickness of $t = 324 \, \mathrm{nm}$ is spin-coated on the wafer as an implantation mask. Subsequently, the sample is baked on a hot plate for $5 \, \mathrm{min}$ at $150^\circ\text{C}$. The nanohole patterns, containing $20 \times20$ of variable diameters $d$ ranging from $30$ to $400 \,  \mathrm{nm}$, were transferred to the photoresist by electron beam lithography (EBL) utilizing a Raith 150TWO system. To tune the number of implanted Si ions through different nanoholes, we vary the nominal nanohole diameter while keeping the EBL dose constant. The overall design including the lateral $10 \, \mathrm{\mu m}$ pitch between all nanoholes was chosen for comparison and consistency with the irradiation pattern used for the FIB writing. During the EBL process, the following parameters are used: $20 \, \mathrm{kV}$ acceleration voltage, $0.25 \, \mathrm{nA}$ current, $30 \, \mathrm{\mu m}$ aperture with a base dose of $820 \, \mathrm{\mu C \cdot cm^{-2}}$. After the EBL, the PMMA is developed with a mixture of DI-water and isopropyl alcohol (3:7) for $30 \, \mathrm{s}$ followed by an isopropyl alcohol  stopper for $30 \, \mathrm{s}$, the samples are then dried with pressurized nitrogen. To create single G centers for VLSI, we use broad-beam implantation with $\mathrm{Si^{2+}}$ ions (energy $40 \, \mathrm{keV}$) through the micro resist mask with a fluence of $\Phi=1 \times 10^{12} \, \mathrm{cm^{-2}}$ at $\theta=$ $7^\circ\text{}$ tilt to avoid ion channeling. After the lift-off process, ultrasonication in acetone for $3 \, \mathrm{min}$ is applied to remove the residuals of PMMA followed by washing in isopropyl alcohol and blow-drying under a stream of nitrogen gas.

According to SRIM calculations \cite{10.1016/j.nimb.2010.02.091}, the $R_p$ of $40 \, \mathrm{keV}$ $\mathrm{Si^{2+}}$ in PMMA is $\sim 100 \, \mathrm{nm}$. Therefore, ions only reach the substrate through the holes in the mask. To prevent the unwanted creation of other types of emitting color centers, no post-irradiation annealing treatment was performed.


\section*{Acknowledgments}
We thank Ilona Skorupa for the help with FA, Gabriele Schnabel for piranha and Bernd Scheumann for associated metal depositions during EBL optimization.  Support from the Ion Beam Center (IBC) at HZDR for ion implantation and Nanofabrication Facilities Rossendorf (NanoFaRo) at IBC is gratefully acknowledged.


\section*{Author contributions}
M.Ho., Y.B and G.V.A. conceived and designed the experiments. M.Ho. performed the single-photon spectroscopy experiments under supervision of G.V.A..  M.Ho., N.K. and L.B. designed the FIB layout. N.K. and L.B. performed FIB implantation and in-situ annealing. M.Ho., N.S.J., Y.B., C.F. and G.V.A. designed the PMMA mask. N.S.J. fabricated the PMMA mask. M.Ho., U.K.,Y.B. and G.V.A.  conceived and performed the broad-beam silicon implantation. C.F. and Y.B. carried out the RTA processing. N.V.A. grew the ultrapure silicon substrates. M.Ho. and G.V.A. wrote the manuscript. All authors, together with G.H., A.E. and M.H. discussed the results and contributed to the manuscript preparation.



\end{document}


\newcommand{\red}[1]{{\color{red}#1}}
\newcommand{\blue}[1]{{\color{blue}#1}}

\renewcommand{\thefigure}{Fig.~S\arabic{figure}}
\renewcommand{\thetable}{S\Roman{table}}
\renewcommand{\figurename}{}
\renewcommand{\theequation}{S\arabic{equation}}



\title{Supplemental Material for \\ 
Wafer-scale nanofabrication of telecom single-photon emitters in silicon}

\author{M.~Hollenbach$^{1,2}$}
\author{N.~Klingner$^{1}$}
\author{N.~S.~Jagtap$^{1,2}$}
\author{L.~Bischoff$^{1}$}
\author{C.~Fowley$^{1}$}
\author{U.~Kentsch$^{1}$}
\author{G.~Hlawacek$^{1}$}
\author{A.~Erbe$^{1}$}
\author{N.~V.~Abrosimov$^{3}$}
\author{M.~Helm$^{1,2}$}
\author{Y.~Berenc\'en$^{1}$}
\author{G.~V.~Astakhov$^{1}$}

\affiliation{$^1$Helmholtz-Zentrum Dresden-Rossendorf, Institute of Ion Beam Physics and Materials Research, 01328 Dresden, Germany  \\
$^2$Technische Universit\"at Dresden, 01062 Dresden, Germany \\ 
$^3$Leibniz-Institut f\"ur Kristallz\"uchtung (IKZ), 12489 Berlin, Germany
 }

\maketitle

\tableofcontents


\section{Experimental setup}

The experimental setup for single-defect spectroscopy is schematically shown in  \ref{S0}. The confocal imaging system is built-up based on a customized closed-cycle helium cryostat (attocube, attoDRY800) that ensures a stable sample temperature of $6.3 \, \mathrm{K}$. The laser light is directed through fiber optics and reflected by a dichroic beamsplitter (DCM1) into a high NA cryo-compatible microscope objective (attocube, LT APO IR, $\mathrm{NA} = 0.81$), resulting in a diffraction-limited focal spot of $1 \, \mathrm{\mu m}$ and a uniform excitation field on the silicon sample plane. For two-dimensional confocal intensity maps, photostability, laser power saturation measurements, and PL spectra, a continuous wave (cw) single-mode 637-nm laser diode (Thorlabs, LP637-SF70) is used. For PL decay measurements, the 785-nm laser light (Thorlabs, LP785-SF20) is pulsed using a fiber-coupled acousto-optical modulator (AA optoelectronic, MT250-NIR6) with a rise time of nominally $6 \, \mathrm{ns}$. For both cw and pulsed excitation, the laser power is controlled using in-line variable fiber optical attenuators (Thorlabs, VOA630-FC/VOA780-FC). To prevent residual BG contribution from the laser lines, the excitation path is cleaned-up with an  800-nm shortpass filter (SP) (Thorlabs, FESH0800). Where quoted, the laser power is measured before the objective and consequently represents an upper limit.

Using three linear nanopositioners (attocube, ANPx311) for precise and independent motion control in the lateral and axial direction ($6 \, \mathrm{mm}$ range and $200 \, \mathrm{nm}$ precision) anchored to the sample holder, three-dimensional confocal PL raster scans are performed. The PL from emitting color centers is collected by the same objective and reflected into the detection path by a second dichroic beamsplitter (DCM2) before being imaged onto a $1 \, \mathrm{mm}$ confocal pinhole (spatial filter). 

A set of $1200 \, \mathrm{nm}$ longpass filters (LP) (Thorlabs, FELH1200) are used to fully remove the residuals of the laser light and unwanted Si bandgap contribution. Depending on the type of the emitter under study, we use a bandpass (BP)  centered either at $1225 \, \mathrm{nm}$ (Edmund Optics, 87-864) or $1275 \, \mathrm{nm}$ (Edmund Optics, 87-865), each with a transmission window of $50 \, \mathrm{nm}$. To spectroscopically separate the superimposed background PL from the emission of the G center in the non-annealed wafer, we use a narrow bandpass (BP) filter with a central wavelength of $1278 \, \mathrm{nm}$ and a nominal  bandwidth of $1 \, \mathrm{nm}$, matching the ZPL position of the G center.

 The emitted photons are detected by a broadband fiber-coupled superconducting nanowire single-photon detector (SNSPD) module (Single Quantum Eos). To measure PL spectra, the signal is redirected into an IR spectrometer equipped with a Peltier-cooled InGaAs photodetector array (PDA)(Andor Technology, iDus $1.7 \, \mathrm{\mu m}$). The SNSPD is used for the spectrally integrated PL measurements and features a detection efficiency of $>90\%$ at $1.3 \, \mathrm{\mu m}$, whereas the PDA provides a quantum efficiency of around $80\%$ in the wavelength range from $1.0 \, \mathrm{\mu m}$ to $1.6 \, \mathrm{\mu m}$. To prove the single-photon nature, we use two SNSPDs in Hanbury Brown and Twiss (HBT) configuration. To this end, we split the PL signal into two detection arms using a 50/50 fiber-optic wideband beamsplitter (Thorlabs, TW1300R5F1).

The lifetime measurements and the photon statistics are recorded with a time-to-digital converter (Swabian Instruments, Time Tagger 20). The synchronization of the measurements is realized by a digital pattern and arbitrary waveform generator (Swabian Instruments, Pulse Streamer 8/2). 

\section{Creation statistics of single G and W centers}

The creation of optically-active point defects or color centers in solids by ion implantation is a stochastic process. In case of single-atom defects, such as the silicon vacancy in SiC or the boron vacancy in hBN, a successful implantation of even a single ion can result in the formation of such color centers, though not every implantation event is successful. The creation yield -- the inverse of the average number of ions required to create a single color center ($\gamma = 1/ \overline{n}$) -- is usually much below $\gamma < 100\%$ despite  ion cascades during implantation. This is because the implantation should create the color center (for instance, vacancy) in a certain charge state, minimally destroy the lattice structure in its vicinity and it should be spatially isolated from other damage-related defects.  The statistical distribution of single-atom color centers can then be well described by the Poisson function  $P_\mu (m) = \mu^m e^{- \mu} / m!$. Here, $m$ is the number of color centers created during the implantation and $\mu$ is the expected value. If the implantation is optimized for the creation of single color centers $\mu = 1$ using  $\overline{n}$ ions, the probability is $P_1 (0) = P_1 (1) = 36.8 \%$. This is also the maximum probability for the Poisson distribution among all possible $\mu$ and $m$.

Our analysis presented in Fig.~4e of the main text shows a probability of $(56 \pm 28)\%$ for the creation of single W centers.  Though within the error bars it falls below the maximum value of $36.8 \%$, there is a strong experimental indication for sub-Poisson distribution of the W centers. We expect this behavior for composite defects, i.e., consisting of more than one atom. For a qualitative explanation, we consider the Poisson distribution of successfully implanted Si ions with  $\mu = 4$ (\ref{S9}a). Under "successful implantation" we understand here the implantation of one over three Si atoms in the tri-interstitial complex forming the W center. Under this assumption, the creation of single and two W centers requires 3 and 6 successful Si implantations, correspondingly. The summation gives probabilities of $23.8\%$, $54.7\%$, $19.5\%$ and $2.0\%$ for zero, single, two and three W centers, respectively (\ref{S9}b). Similar arguments can be applied for the formation of the G center, consisting of two C atoms and one Si atom. The real formation processes of the W and G centers under Si implantation are much more complex than in the simplified consideration above and beyond the scope of this work. Nevertheless, our approach for the calculation of sub-Poisson distribution reproduces well experiments in Fig.~2b and Fig.~4e with $\mu = 4$ and $\mu = 4.8$, respectively.


\newpage

\section{Supporting tables}

\begin{table}[h!]
\caption{Average number of implanted Si ions per spot in the FIB pattern. } 
\begin{tabular}{c | c}
Row & Average Si ions per spot   \\ \hline \hline
1 & 6  \\ \hline
2 & 9  \\ \hline
3 & 13  \\ \hline
4 & 16  \\ \hline
5 & 25  \\ \hline
6 & 33  \\ \hline
7 & 45  \\ \hline
8 & 61  \\ \hline
9 & 83  \\ \hline
10 & 113  \\ \hline
11 & 153  \\ \hline
12 & 208  \\ \hline
13 & 283  \\ \hline
14 & 384  \\ \hline
15 & 500   
\end{tabular}
\end{table}

\begin{table}[h!]
\caption{Variation of the nanohole diameter in the PMMA pattern} 
\begin{tabular}{c | c}
Row & Nominal hole diameter (nm)   \\ \hline \hline
1 & 30  \\ \hline
2 & 35  \\ \hline
3 & 40  \\ \hline
4 &  45 \\ \hline
5 &  50 \\ \hline
6 &   55 \\ \hline
7 &   60 \\ \hline
8 &  65 \\ \hline
9 &   70 \\ \hline
10 &  75 \\ \hline
11 &  80 \\ \hline
12 &  85 \\ \hline
13 &  90 \\ \hline
14 &  95 \\ \hline
15 &  100 \\ \hline
16 &  125 \\ \hline
17 &  150 \\ \hline
18 &  200 \\ \hline
19 &  300 \\ \hline
20 &  400 \\ \hline
21 & 2000
\end{tabular}
\end{table}

\clearpage

\section{Supporting figures}

\begin{figure}[h!]
\centerline{\includegraphics[width=0.89\columnwidth,clip]{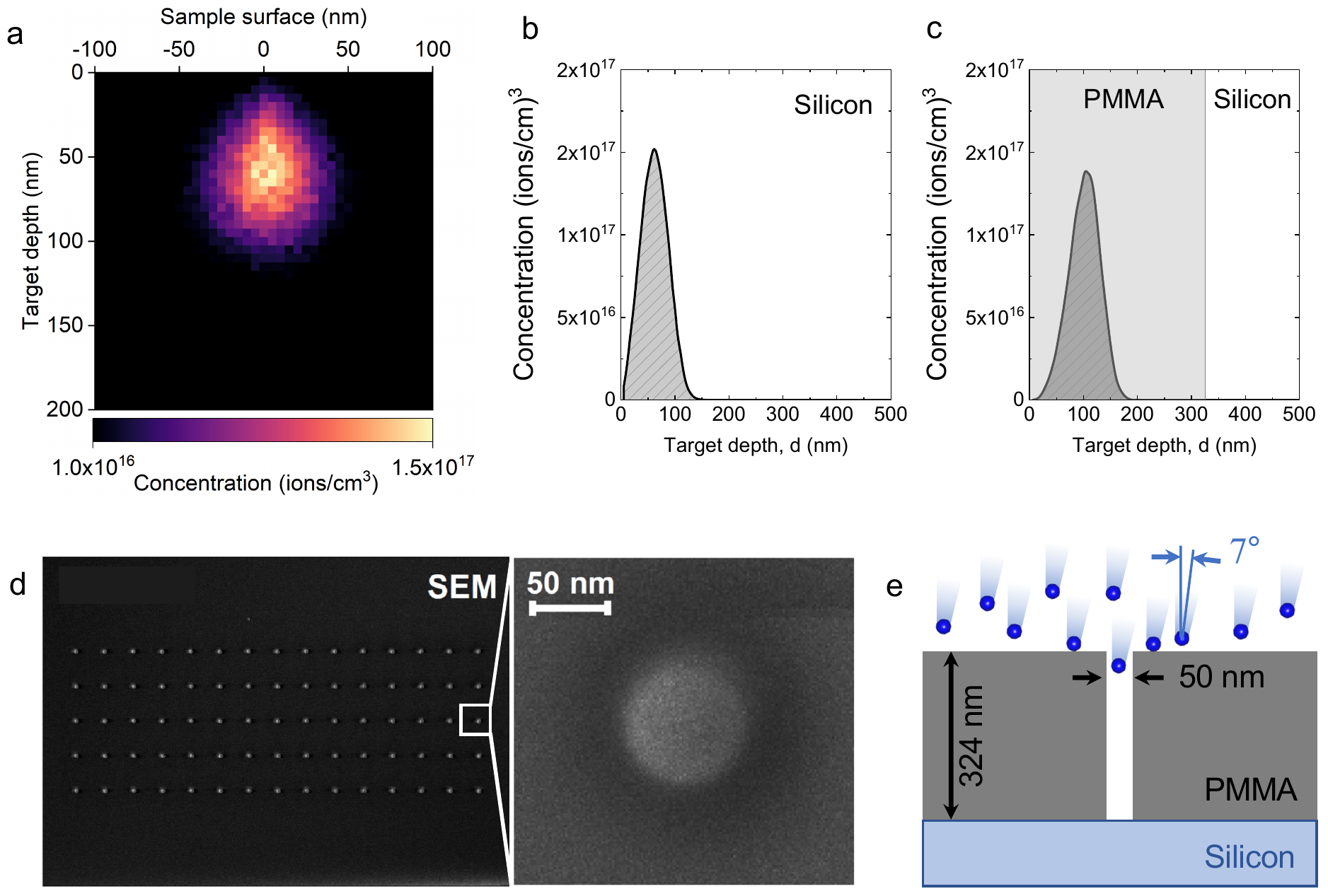}}
\caption{a, Spatially resolved SRIM simulation of the implantation with double charged Si$^{2+}$ ions with a landing energy of $40 \, \mathrm{keV}$ into the Si lattice indicating longitudinal and lateral straggling. b, SRIM simulation of the Si implantation stopping range in Si lattice ($R_p \sim$ 60 nm). c, SRIM simulation of the Si implantation stopping range in PMMA ($R_p \sim$ 100 nm). d, SEM image of a sample subset of nanohole arrays patterned on the PMMA surface. Zoom-in shows an SEM micrograph of an aperture with a nominal diameter of approximately $d = 50 \, \mathrm{nm}$. e, Schematic cross-section of the Si implantation through a nanohole in the PMMA mask with a nominal thickness $t = 324 \, \mathrm{nm}$. Implantation is performed at $\theta = 7 ^{\circ}$ tilt to avoid ion channeling. } \label{S5}
\end{figure}

\begin{figure}[h!]
\centerline{\includegraphics[width=0.93\columnwidth,clip]{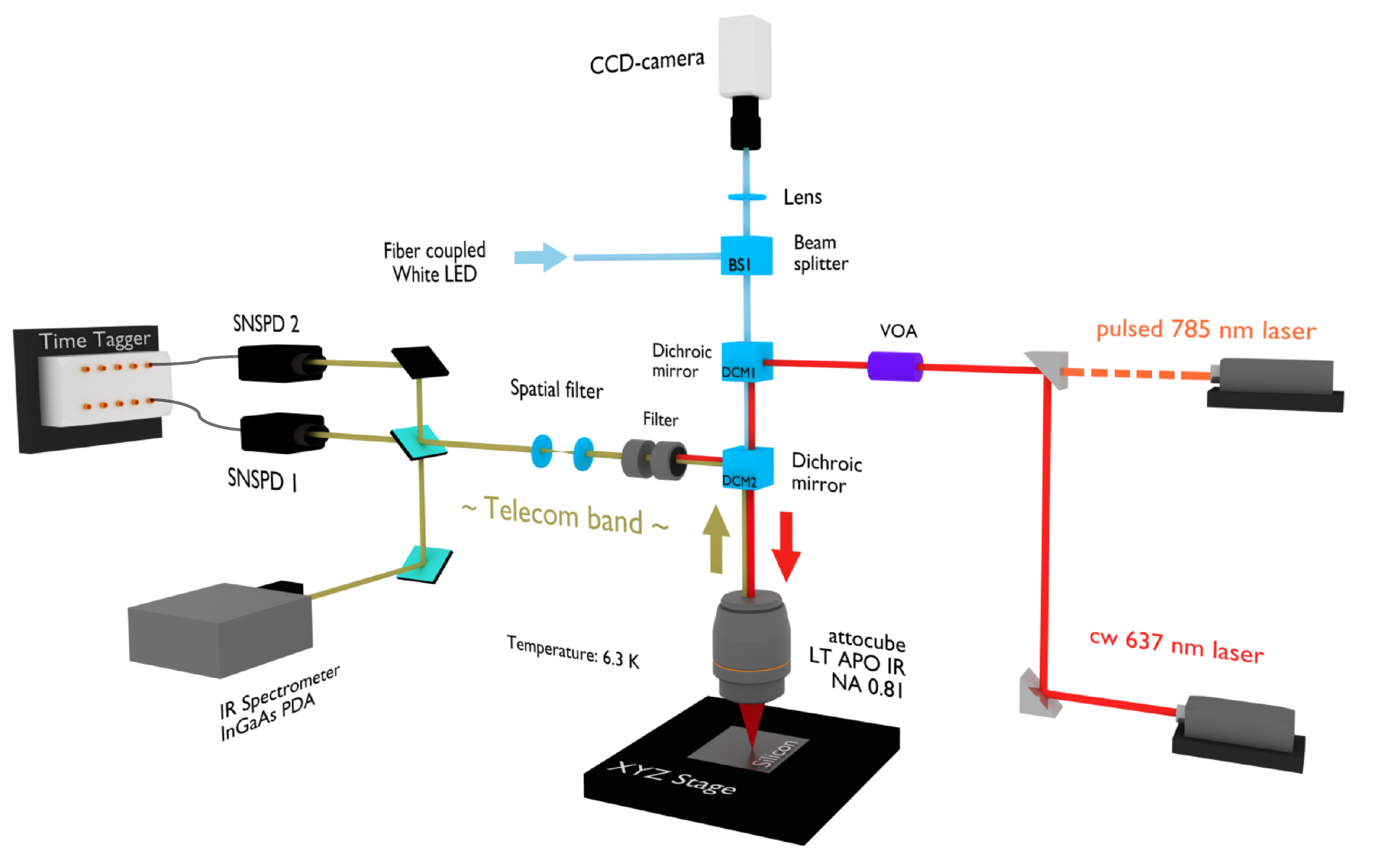}}
\caption{A scheme of the experimental setup. The PL is excited with either cw 637-nm or pulsed 785-nm laser. The PL is collected through optical LP and BP filters followed by a spatial filter. The detection is performed either with two SNSPDs or an InGaAs PDA. } \label{S0}
\end{figure}

\begin{figure}[h!]
\centerline{\includegraphics[width=0.97\columnwidth,clip]{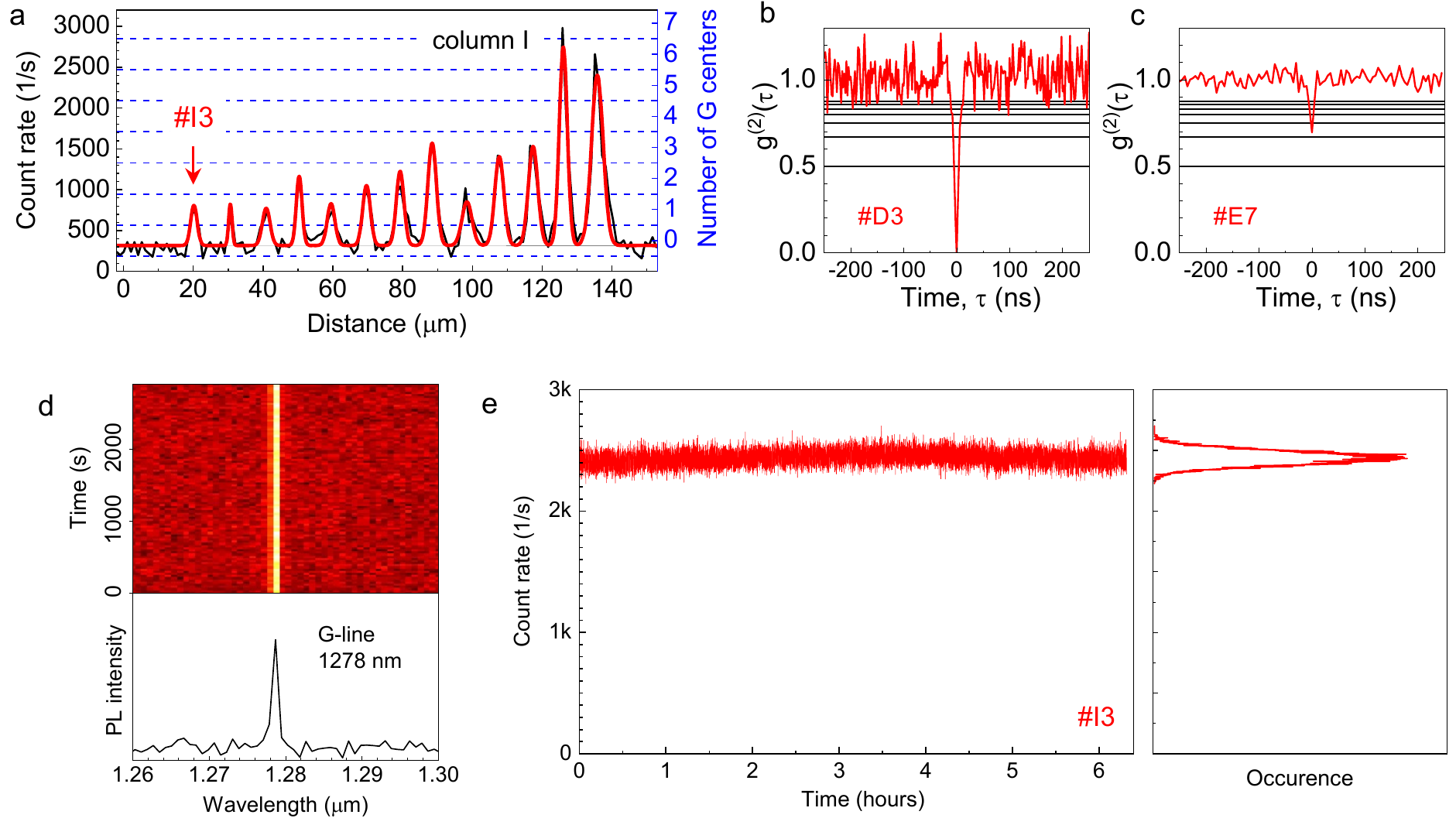}}
\caption{a, G center ZPL line scan through column I of the FIB pattern in Fig.~1 of the main text.  The red line is a multi-Gauss fit. The right axis represents the number of G centers ($N$) according to the count rate. b, Second-order autocorrelation function $g^{(2)} (\tau)$ obtained at spot  $\#$D3 with BG correction, yielding $N = 1$. c, Second-order correlation function $g^{(2)} (\tau)$ obtained at spot  $\#$E7 with BG correction, yielding $N = 3$. d, ZPL time stability of a single G center obtained with a 1-nm BP filter about one hour. e, PL time trace of a single G center without  BP filter for over 6 hours. } \label{S1}
\end{figure}

\begin{figure}[h!]
\centerline{\includegraphics[width=0.89\columnwidth,clip]{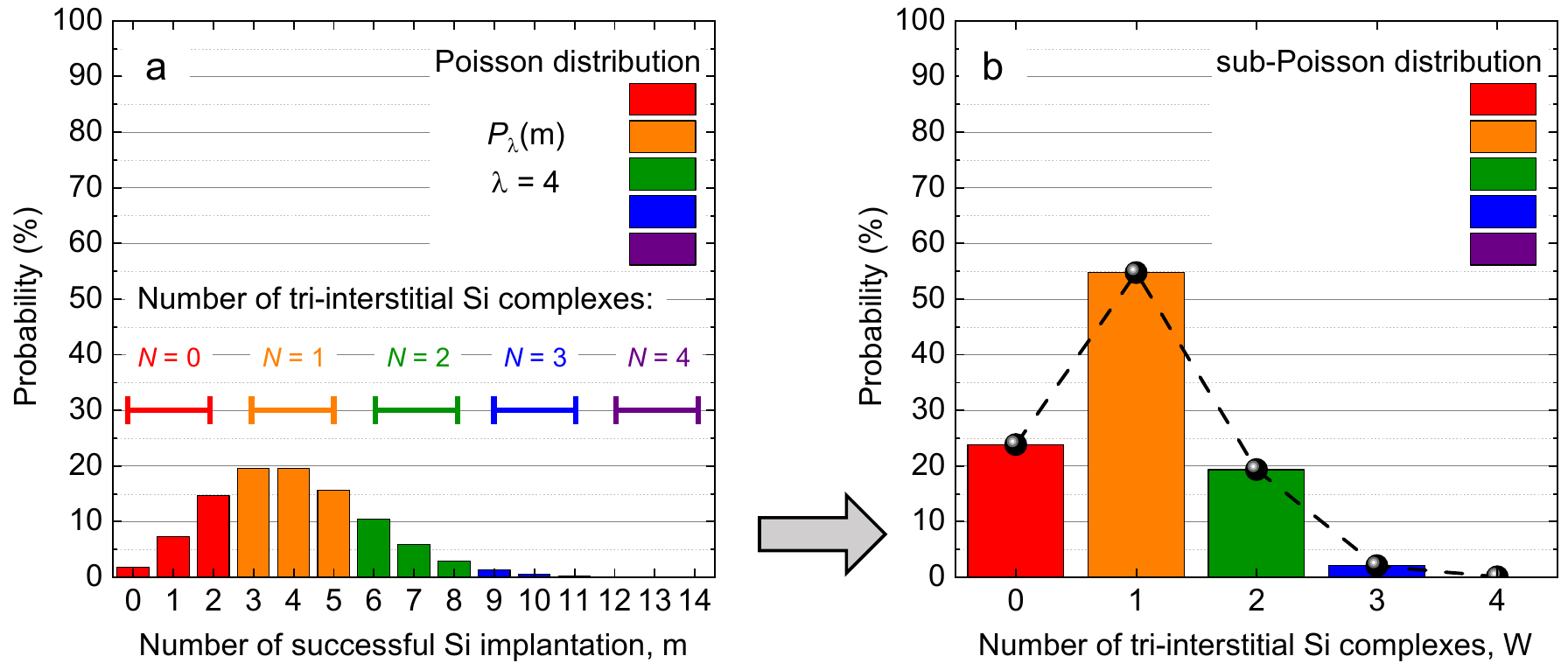}}
\caption{a, Poisson distribution $P_\mu (m)$ of successfully implanted Si ions with expected value $\mu = 4$ per each implantation. b,  Corresponding distribution of tri-interstitial Si complexes with sub-Poission statistics.} \label{S9}
\end{figure}

\begin{figure}[h!]
\centerline{\includegraphics[width=0.79\columnwidth,clip]{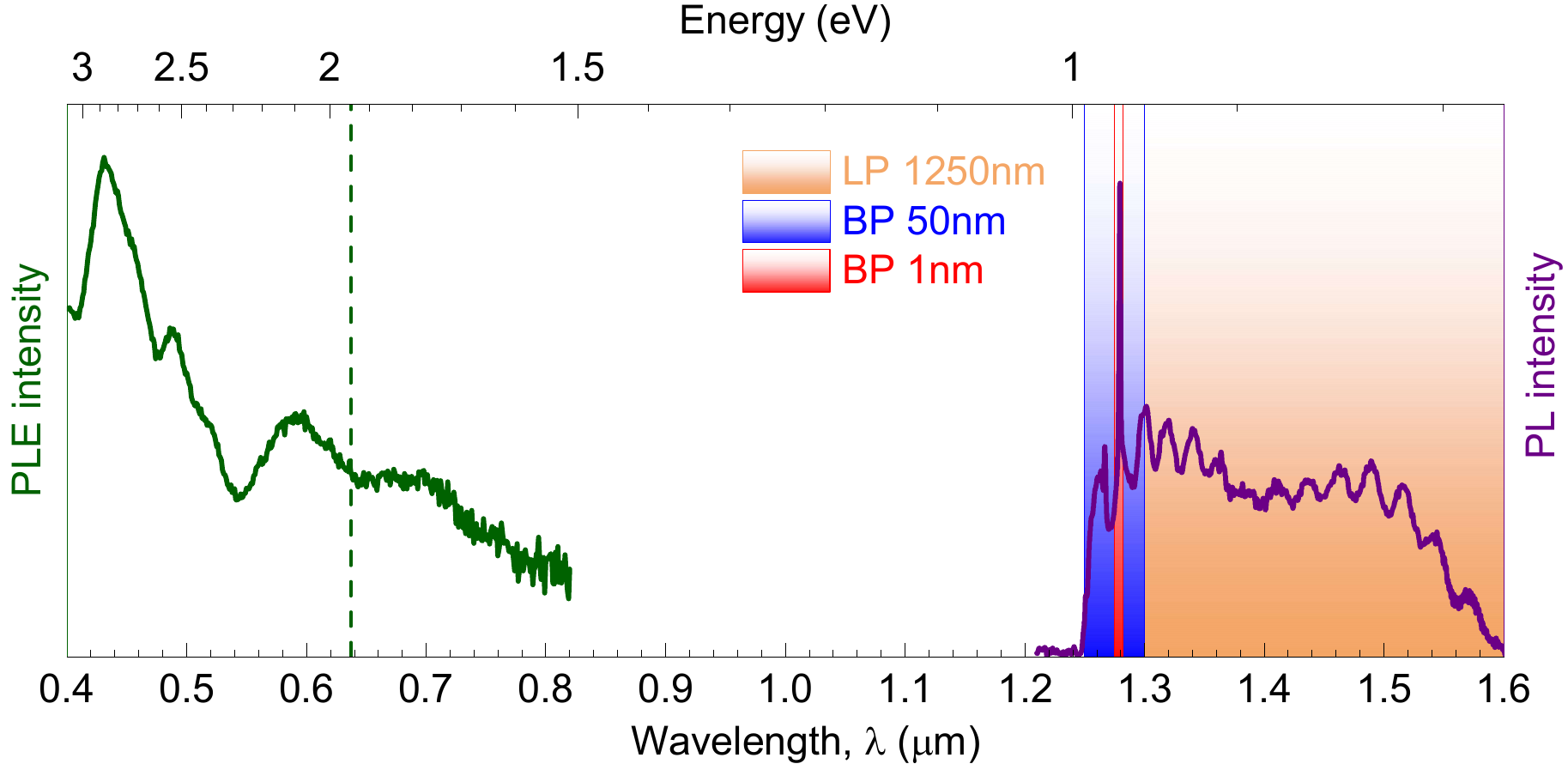}}
\caption{PL spectrum of a single G center before annealing and PLE spectrum of the G center ensemble created by broad-beam implantation. The PLE intenisty is monitored at the ZPL position $\lambda_G =1278 \, \mathrm{nm}$. The vertical dashed line indicates the excitation wavelength $\lambda_{exc} = 637 \, \mathrm{nm}$. }\label{S3}
\end{figure}

\begin{figure}[h!]
\centerline{\includegraphics[width=0.83\columnwidth,clip]{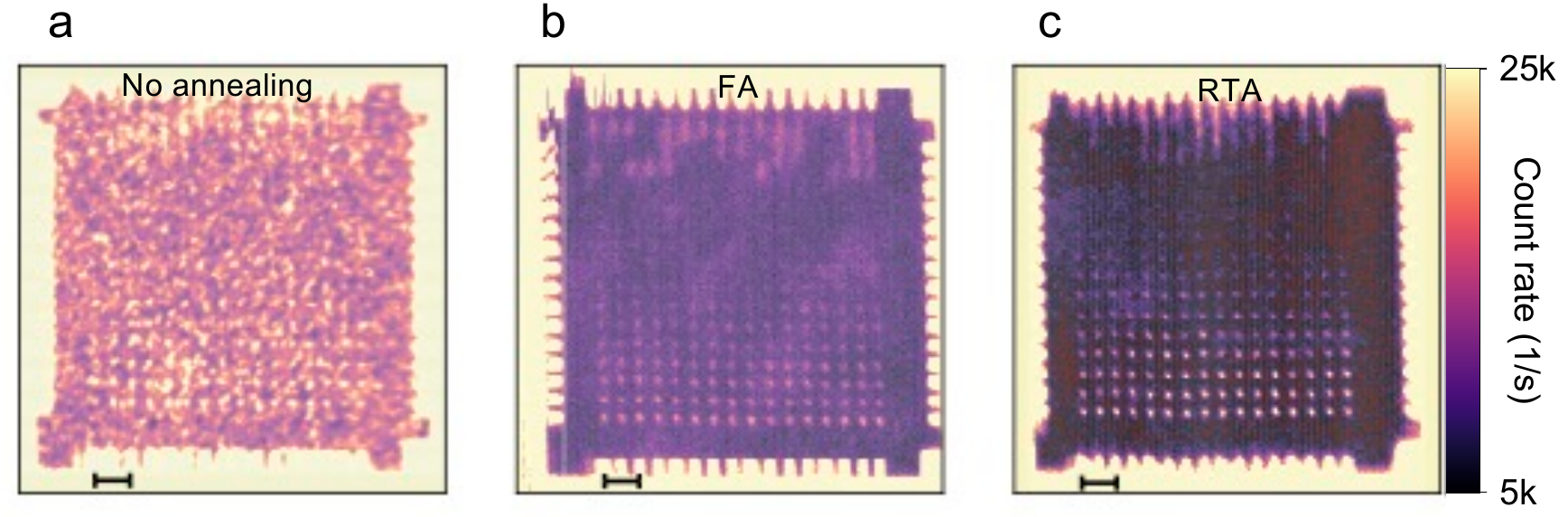}}
\caption{a, Confocal PL intensity map as  in Fig.~1 of the main text but measured with a long-pass filter $\lambda > 1250 \, \mathrm{nm}$ without annealing. b, Confocal PL intensity map after furnace annealing (FA) at $500^{\circ}\mathrm{C}$ over 2 hours in an $\mathrm{N_{2}}$ atmosphere. The FIB pattern is created after annealing. c, Confocal PL intensity map after rapid thermal annealing (RTA) at $1000^{\circ}\mathrm{C}$ over 60 seconds in an $\mathrm{N_{2}}$ atmosphere. The FIB pattern is created after annealing. The scale bar in a-c  is $20 \, \mathrm{\mu m}$. } \label{S2}
\end{figure}

\begin{figure}[h!]
\centerline{\includegraphics[width=0.79\columnwidth,clip]{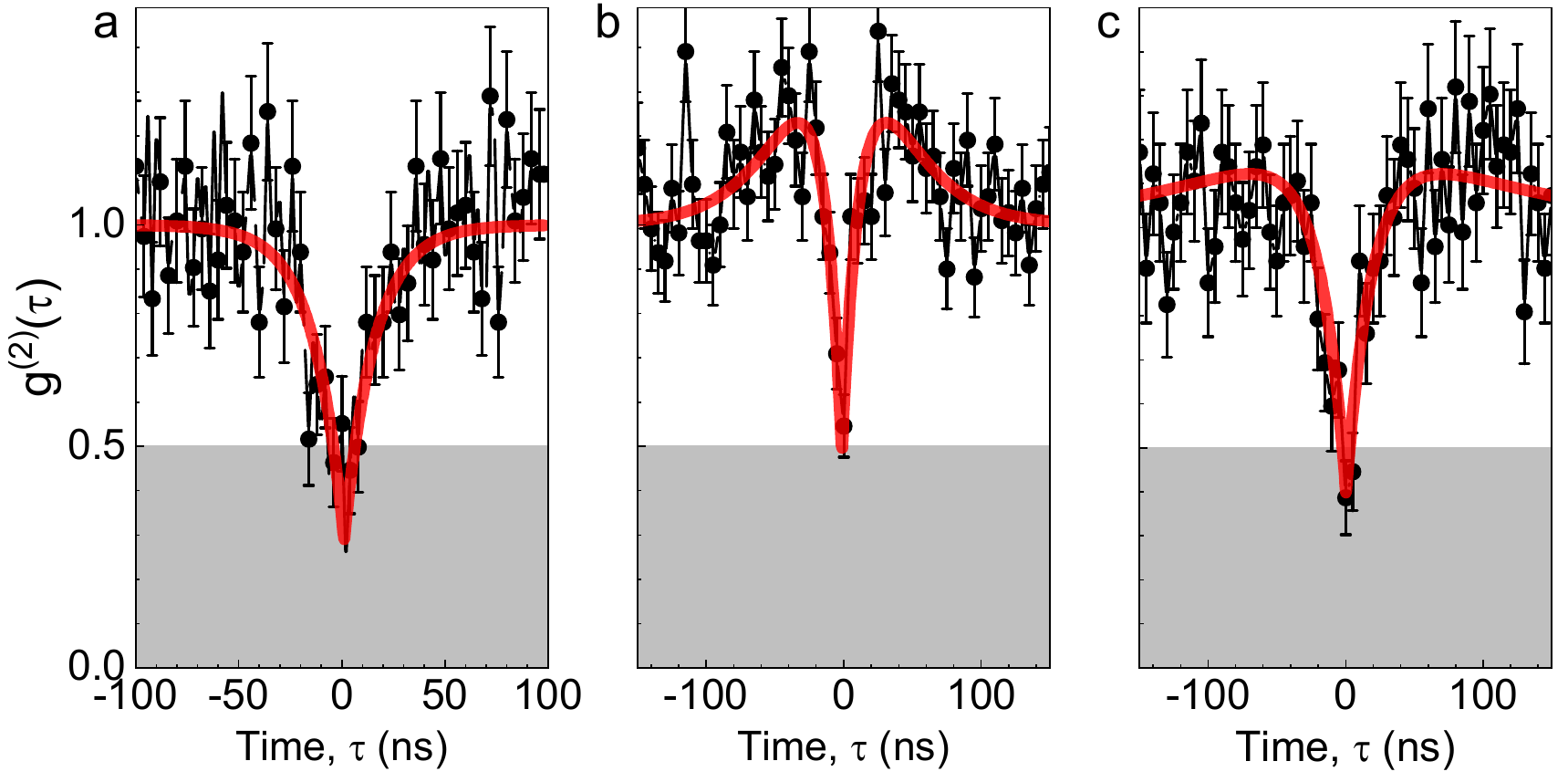}}
\caption{Examples of the second-order autocorrelation functions $g^{(2)} (\tau)$ obtained in an annealed SOI wafer after broad-beam Si implantation through PMMA nanoholes (Fig.~2 of the main text). No BG corrections are performed. In all cases, the number of G centers is $N = 1$.} \label{S4}
\end{figure}

\begin{figure}[h!]
\centerline{\includegraphics[width=0.79\columnwidth,clip]{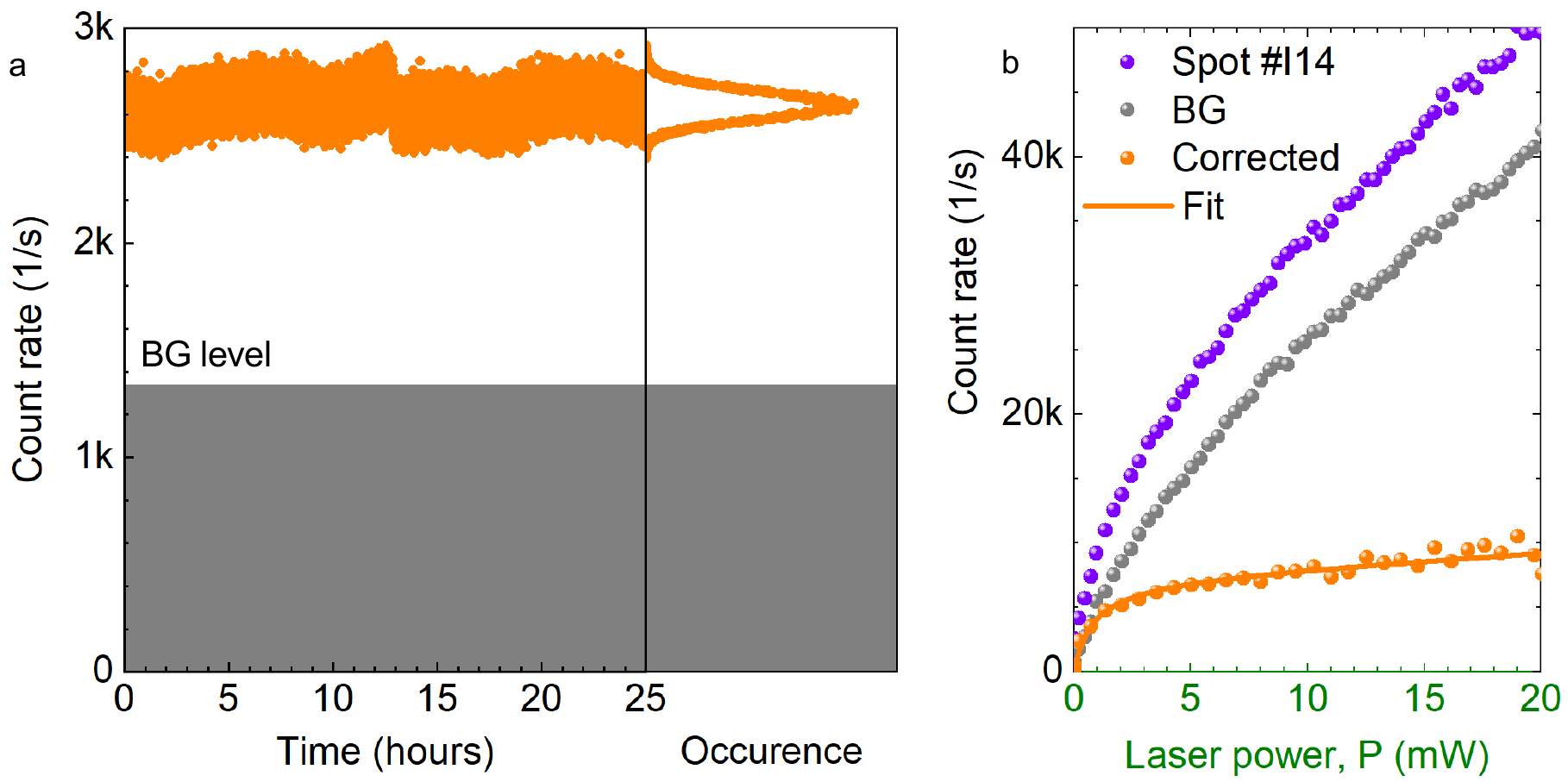}}
\caption{a, PL time trace of a single W center (spot \#I14) with a BP filter $\Delta \lambda = 50 \, \mathrm{nm}$. b,  Count rate of a single W center as a function of the excitation power in the presence of BG. The power dependence of the BG is also shown for comparison.  The corrected power dependence corresponds to Fig.~4 of the main text. } \label{S7}
\end{figure}

\begin{figure}[h!]
\centerline{\includegraphics[width=0.79\columnwidth,clip]{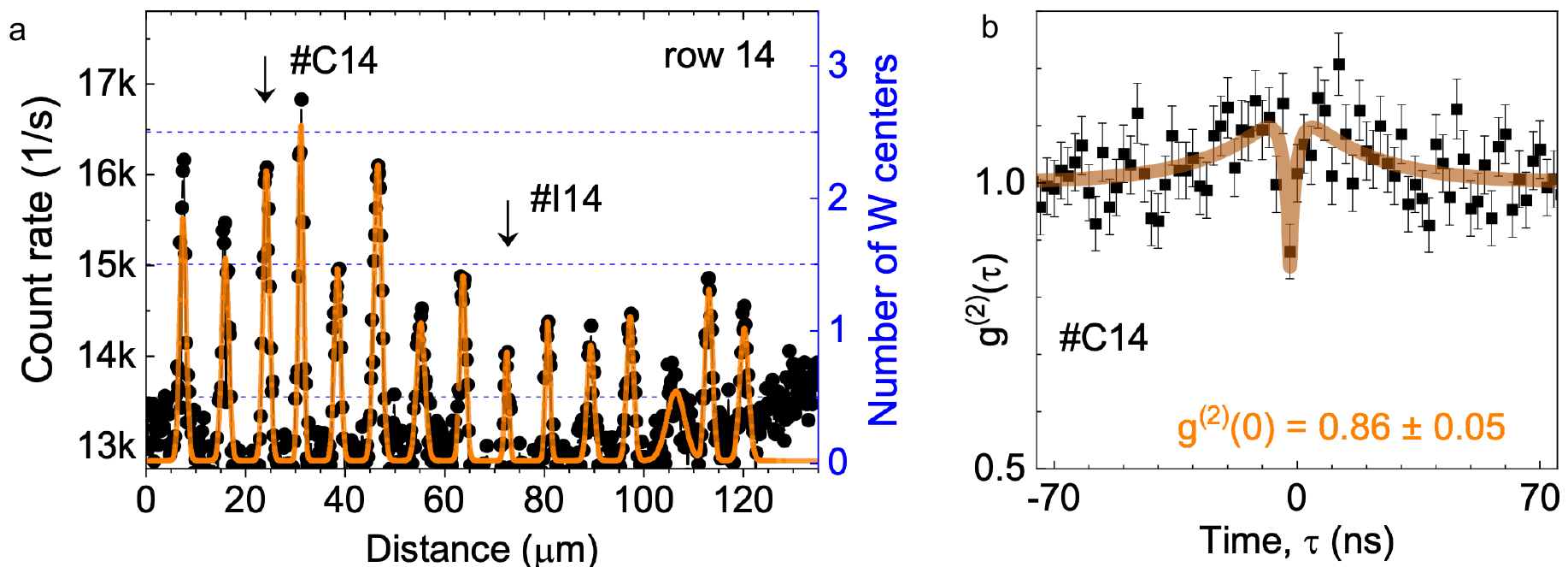}}
\caption{a, W center PL line scan through the line 14 of the FIB pattern in Fig.~4 of the main text. The orange line is a multi-Gauss fit. The right axis represents the number of W centers ($N$) assuming the count rate from each spot to be proportinal to the number of W-centers within this spot. b,  Second-order autocorrelation function $g^{(2)} (\tau)$ obtained at spot $\#$C14 without any BG correction. The BG correction gives $g_{corr}^{(2)} (0) = 0.52 \pm 0.15$, corresponding two $N = 2$. } \label{S6}
\end{figure}